\documentclass[]{spie}  
\usepackage[]{graphicx}
\usepackage{aas_macro}
\usepackage{amsmath,amssymb,amsbsy,mathtools,url}
\title{\Large \bf Review of high-contrast imaging systems for current and future ground- and space-based telescopes\\
\large I. Coronagraph design methods and optical performance metrics} 
\usepackage{array}
\usepackage{hyperref}
\newcolumntype{P}[1]{>{\centering\arraybackslash}p{#1}}

\author{
\large
G.~Ruane\supit{a,$\dagger$}, A.~Riggs\supit{b}, J.~Mazoyer\supit{c}, E.~H.~Por\supit{d}, M.~N'Diaye\supit{e}, E.~Huby\supit{f}, P.~Baudoz\supit{f}, R.~Galicher\supit{f}, E.~Douglas\supit{g}, J.~Knight\supit{h}, B.~Carlomagno\supit{i}, K.~Fogarty\supit{j}, L.~Pueyo\supit{j}, N.~Zimmerman\supit{k}, O.~Absil\supit{i}, M.~Beaulieu\supit{e}, E.~Cady\supit{b}, A.~Carlotti\supit{l}, D.~Doelman\supit{d}, O.~Guyon\supit{h,m,n}, S.~Haffert\supit{d}, J.~Jewell\supit{b}, N.~Jovanovic\supit{a}, C.~Keller\supit{d}, M.~A.~Kenworthy\supit{d},  J.~K\"{u}hn\supit{o}, K.~Miller\supit{h}, D.~Sirbu\supit{p}, F.~Snik\supit{d}, J.~Kent~Wallace\supit{b}, M.~Wilby\supit{d}, and M.~Ygouf\supit{q} \\
\small
\supit{a}Department of Astronomy, California Institute of Technology, 1200 E. California Blvd., Pasadena, CA 91125, USA \\
\supit{b}Jet Propulsion Laboratory, California Institute of Technology, 4800 Oak Grove Dr., Pasadena, CA 91109, USA\\
\supit{c}Johns Hopkins University, Zanvyl Krieger School of Arts and Sciences, Department of Physics and Astronomy, Bloomberg Center for Physics and Astronomy, 3400 North Charles Street, Baltimore, MD 21218, USA\\
\supit{d}Leiden Observatory, Leiden University, P.O. Box 9513, 2300 RA Leiden, The Netherlands\\
\supit{e}Universit\'{e} C\^{o}te d'Azur, Observatoire de la C\^{o}te d'Azur, CNRS, Laboratoire Lagrange, Bd de l'Observatoire, CS 34229, 06304 Nice cedex 4, France\\
\supit{f}LESIA, Observatoire de Paris, PSL Research University, CNRS, Sorbonne Universités, Univ. Paris Diderot, UPMC Univ. Paris 06, Sorbonne Paris Cit\'{e}, 5 place Jules Janssen, 92190 Meudon, France\\
\supit{g}Department of Aeronautics and Astronautics, Massachusetts Institute of Technology, Cambridge, MA 02139, USA\\
\supit{h}University of Arizona, Tucson, AZ 85721, USA\\
\supit{i}Space sciences, Technologies, and Astrophysics Research (STAR) Institute, University of Li\`{e}ge, Li\`{e}ge, Belgium\\
\supit{j}Space Telescope Science Institute, 3700 San Martin Dr, Baltimore, MD 21218, USA\\
\supit{k}NASA Goddard Space Flight Center, 8800 Greenbelt Rd., Greenbelt, MD, 20771 USA\\
\supit{l}Institut de Planétologie et d'Astrophysique de Grenoble (IPAG), Grenoble, France \\
\supit{m}Astrobiology Center, National Institutes of Natural Sciences, 2-21-1 Osawa, Mitaka, Tokyo, Japan\\
\supit{n}National Astronomical Observatory of Japan, Subaru Telescope, National Institutes of Natural Sciences, Hilo, HI 96720, USA\\
\supit{o}Institute for Particle Physics and Astrophysics, ETH Zurich, 8093 Switzerland\\
\supit{p}NASA Ames Research Center, Moffett Field, Mountain View, CA, 94035, USA\\
\supit{q}IPAC, California Institute of Technology, 1200 E. California Blvd., Pasadena, CA 91125, USA
}

\authorinfo{$^\dagger$Send correspondence to gruane@caltech.edu}

\graphicspath{{Figures/}}
 
\begin{document} 
  \maketitle 

\begin{abstract}
The Optimal Optical Coronagraph (OOC) Workshop at the Lorentz Center in September 2017 in Leiden, the Netherlands gathered a diverse group of 25 researchers working on exoplanet instrumentation to stimulate the emergence and sharing of new ideas. In this first installment of a series of three papers summarizing the outcomes of the OOC workshop, we present an overview of design methods and optical performance metrics developed for coronagraph instruments. The design and optimization of coronagraphs for future telescopes has progressed rapidly over the past several years in the context of space mission studies for Exo-C, WFIRST, HabEx, and LUVOIR as well as ground-based telescopes. Design tools have been developed at several institutions to optimize a variety of coronagraph mask types. We aim to give a broad overview of the approaches used, examples of their utility, and provide the optimization tools to the community. Though it is clear that the basic function of coronagraphs is to suppress starlight while maintaining light from off-axis sources, our community lacks a general set of standard performance metrics that apply to both detecting and characterizing exoplanets. The attendees of the OOC workshop agreed that it would benefit our community to clearly define quantities for comparing the performance of coronagraph designs and systems. Therefore, we also present a set of metrics that may be applied to theoretical designs, testbeds, and deployed instruments. We show how these quantities may be used to easily relate the basic properties of the optical instrument to the detection significance of the given point source in the presence of realistic noise. 

\end{abstract}


\keywords{High contrast imaging, instrumentation, exoplanets, direct detection, coronagraphs}

\section{INTRODUCTION}
\label{sec:intro}  

The field of coronagraph instrument design has advanced considerably since Bernard Lyot initially inserted a brass disk in the focal plane of his instrument to take the first images of the solar corona without an eclipse circa 1930\cite{Lyot1939}. In recent decades, coronagraphy has expanded well beyond its original purpose. The idea to use Lyot's coronagraph to observe the circumstellar environment of nearby stars came in the 1970's \cite{Bonneau1975}, leading to the detection of the first debris disks in the 1980's and 1990's \cite{Smith1984,Golimowski1993,Kalas1995,Mouillet1997}. With major improvements to observatories, the advent of adaptive optics (AO) in astronomy\cite{Babcock1953,Beckers1993,Troy2000,Oppenheimer2001}, and the discovery of exoplanets \cite{Campbell1988,Wolszczan1992,Mayor1995}, coronagraphs are now routinely used at the World's largest telescope facilities to peer within a fraction of arcsecond of stars to search for and characterize faint companions, including brown dwarfs\cite{Nakajima1995,Oppenheimer2001} and giant exoplanets\cite{Marois2008,Lagrange2009,Macintosh2015}, and map the dust distributions of circumstellar disks. It is also becoming increasingly feasible to discover and characterize potentially habitable planets and detect potential biomarkers via direct imaging and spectroscopy with current and future ground- and space-based telescopes \cite{Davies1980,Trauger2007}. 

The Optimal Optical Coronagraph (OOC) Workshop at the Lorentz Center in September 2017 in Leiden, the Netherlands gathered 25 researchers to facilitate progress and international collaboration in the field of instrumentation for exoplanet imaging. In this first installment of a series of three papers summarizing the outcomes of the OOC workshop, we review optical performance metrics and optimization methods developed for coronagraph instruments.


\section{Performance metrics}

In this section, we define the key performance metrics considered during the design of a coronagraph instrument for ground- and space-based telescopes. 

The best feasible combination of performance parameters is the one that ultimately maximizes the estimated scientific yield of an instrument or mission. Depending on the observing goals or circumstances (e.g., wavelength coverage and desired field of view), more than one coronagraph may be used in a single instrument to maximize multiple yield metrics, including the number of planets discovered and characterized. Optimizing the coronagraph masks to do so requires a thorough understanding of the dominant noise sources present in the system.

In the case of ground-based telescopes, coronagraphs are generally situated downstream of a closed-loop AO system that corrects for atmospheric turbulence. Coronagraph design for ground-based telescopes relies on an analysis of the expected wavefront errors (as in e.g. Fusco et al. 2006 \cite{Fusco2006}). Coronagraph designs therefore evolve with the rapid development of deformable mirror (DM) and wavefront sensing technology, which are pushing the performance of adaptive optics toward the conventionally accepted limits of AO systems \cite{Guyon2005_AO} and potentially even beyond that using predictive control \cite{Males2018}. 

Although space telescopes have the advantage of being above the Earth's atmosphere, wavefront sensing and control is still a critical aspect of space-based exoplanet imaging with coronagraphs. The goal of imaging Earth-like planets in the habitable zone around solar type stars drives wavefront correction and stability to picometers of error at mid-spatial frequencies (approximately 2-30 cycles per pupil diameter) to maintain small enough levels of leaked starlight. Other dominant noise sources include detector read noise and dark current.

\subsection{Engineering metrics}

Many papers have defined metrics that describe the optical performance of coronagraphs\cite{Guyon2005,Krist2015,Nemati2017,JensenClem2018}. Here, we outline a number of useful metrics and relate them to the smallest detectable planet-to-star flux ratio. 

\begin{figure}[t]
    \centering
    \includegraphics[width=0.7\linewidth]{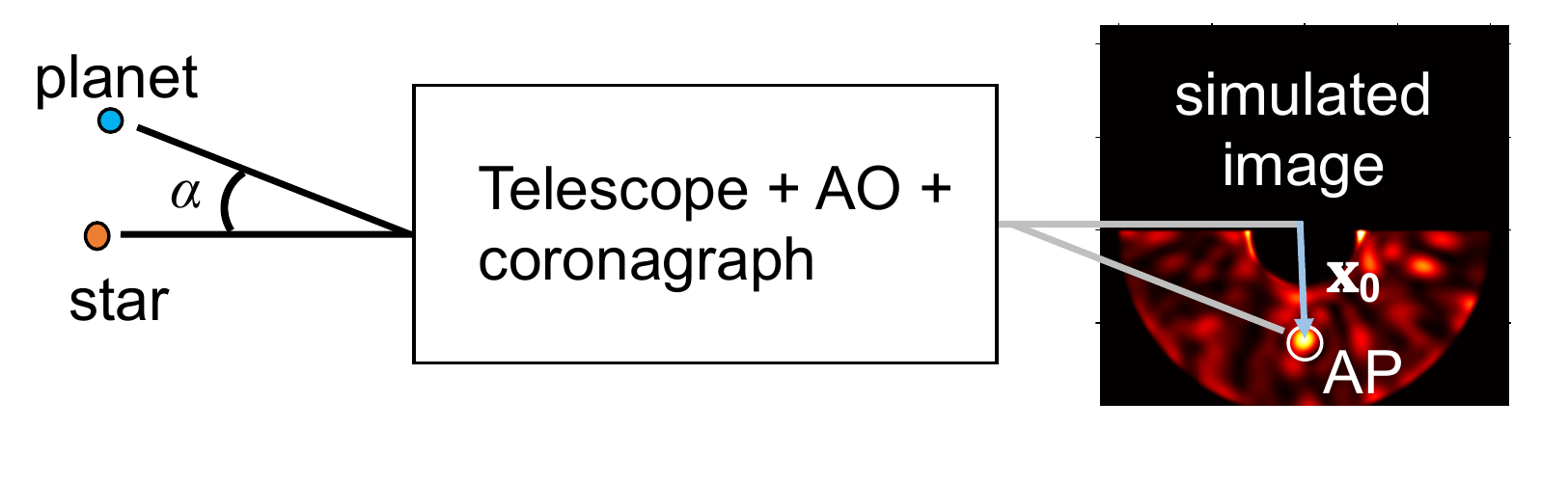}
    \caption{Qualitative schematic of a system to directly image exoplanets. The star and planet are separated by angle $\alpha$. The star is aligned to the optical axis and suppressed by the coronagraph. The planet image appears at a distance $\mathbf{x_0}$ from the optical axis. The signal from the planet is integrated within the aperture for photometric estimation (AP). The diffracted starlight that appears within AP is a dominant source of noise. }
    \label{fig:layout}
\end{figure}

We define two types of point spread functions (PSFs). The first, $\text{PSF}_0(\mathbf{x},\lambda)$, is the PSF of the imaging system assuming a perfect wavefront, flat deformable mirrors, and all of the coronagraph masks removed including apodizers, focal plane masks, and Lyot stops. $\mathbf{x}$ is the position vector in the image plane and $\lambda$ is the wavelength. In this case, the PSF is theoretically shift invariant. The second type, $\text{PSF}_\text{coro}(\mathbf{x},\mathbf{x_0},\lambda)$, is the PSF with the coronagraph masks in place and DMs set to minimize starlight in the desired ``dark hole" region of the image plane. $\text{PSF}_\text{coro}$ is different from $\text{PSF}_0$ in that it is varies as a function of source position $\mathbf{x_0}$. The PSFs are normalized such that 
\begin{equation}
    \int \text{PSF}_0(\mathbf{x}) d\mathbf{x} = 1
\end{equation}
and the total energy in $\text{PSF}_\text{coro}$ is the fraction of light that passes through the coronagraph masks.

The signal detected in photo-electrons from the planet and star at position $\mathbf{x_0}$ is
\begin{align}
    S_p(\mathbf{x_0}) &= \int_{\Delta\lambda} \eta_p(\mathbf{x_0},\lambda) \Phi_p(\lambda) A \Delta t \, q(\lambda) T(\lambda)d\lambda,\\
    S_s(\mathbf{x_0}) &= \int_{\Delta\lambda} \eta_s(\mathbf{x_0},\lambda) \Phi_s(\lambda) A \Delta t \, q(\lambda) T(\lambda)d\lambda,
\end{align}
respectively, where $\eta_p$ and $\eta_s$ are the fraction of collected planet and star light that is detected. Specifically, $\eta_p$ and $\eta_s$ may be computed by
\begin{equation}
   \eta_p(\mathbf{x_0},\lambda) = \int_{\text{AP}(\mathbf{x_0})} \text{PSF}_\text{coro}(\mathbf{x},\mathbf{x_0},\lambda) d\mathbf{x}
\end{equation}
\begin{equation}
   \eta_s(\mathbf{x_0},\lambda) = \int_{\text{AP}(\mathbf{x_0})} \text{PSF}_\text{coro}(\mathbf{x},0,\lambda) d\mathbf{x},
\end{equation}
where each is integrated over a finite circular aperture for photometric estimation in the image plane, $\text{AP}(\mathbf{x_0})$, centered on $\mathbf{x_0}$. Alternatively, $\eta_p$ and $\eta_s$ may also be defined in terms of the result of a matched-filter or the fraction of light coupled into single mode fibers, for example. $\Phi_p(\lambda)$ and $\Phi_s(\lambda)$ are the flux owing to the planet and star (photons per unit area per unit time per unit wavelength at the primary mirror), $\Delta t$ is the integration time, $\Delta\lambda$ is the spectral bandwidth, $A$ is the collecting area of the telescope, $q(\lambda)$ is the detective quantum efficiency, and $T(\lambda)$ is the transmission of the instrument describing losses that affect the star and planet equally. The variables are summarized in Table \ref{table:vars}.

The most widely used coronagraph performance metric is raw contrast defined by 
\begin{equation}
    C(\mathbf{x_0},\lambda) = \frac{\eta_s(\mathbf{x_0},\lambda)}{\eta_p(\mathbf{x_0},\lambda)}.
\end{equation}
The raw contrast may also be integrated in over spatial and spectral dimensions. 

The throughput is defined in a number of ways. The absolute coronagraph throughput is $\eta_p(\mathbf{x_0},\lambda)$. The relative throughput is normalized to the throughput of the system without the coronagraph masks: $\eta_p(\mathbf{x_0},\lambda)/\eta_\text{tel}(\lambda)$, where 
\begin{equation}
    \eta_\text{tel}(\lambda) = \int_{\text{AP}(0)} \text{PSF}_0(\mathbf{x},\lambda) d\mathbf{x}.
\end{equation}
The total energy throughput is the integral of $\eta_p(\mathbf{x_0},\lambda)$ with respect to $\mathbf{x_0}$ over the image plane and with respect to $\lambda$ over the spectral bandwidth $\Delta\lambda$. The spectral bandwidth is most often reported as a fraction ($\Delta\lambda/\lambda$) or percentage ($\Delta\lambda/\lambda\times100$) for broadband imaging, whereas the spectral resolution of spectrographs is typically given by $R=\lambda/\Delta\lambda$.

\begin{table}[t!]
\small
\centering
\begin{tabular}{|l|l|} 
 \hline
Parameter & Definition\\
\hline
$\text{PSF}_0$ & Point spread function with all coronagraph masks removed. Shift invariant.\\
\hline
$\text{PSF}_\text{coro}$ & Point spread function with the coronagraph masks in place. Shift variant.\\
\hline
$S_p$ & Detected signal from the planet (photo-electrons).\\
\hline
$S_s$ & Detected signal from the star at the position of the planet (photo-electrons).\\
\hline
$\eta_p$ & Fraction of available planet light detected. The ``absolute coronagraph throughput."\\
\hline
$\eta_s$ & Fraction of available star light detected. \\
\hline
$\eta_\text{tel}$ & Fraction of light detection from a point source with coronagraph masks removed. \\
\hline
$\Phi_p$ & Planet flux (photons/m$^2$/$\mu$m/sec at the primary mirror).\\
\hline
$\Phi_s$ & Stellar flux (photons/m$^2$/$\mu$m/sec at the primary mirror).\\
\hline
$A$ & Area of the telescope aperture (m).\\
\hline
$\Delta t$ & Integration time (sec).\\
\hline
 $q$ & Quantum efficiency. \\
\hline
$T$ & Telescope transmission. Accounts for losses that affect the planet and star equally.\\
\hline
$\lambda$ & Wavelength ($\mu$m).\\
\hline
$\Delta\lambda$ & Spectral bandwidth ($\mu$m).\\
\hline
$R$ & Spectral resolution.\\
\hline
AP & Aperture for photometric estimation.\\
\hline
$C$ & $C=\eta_s/\eta_p$. Raw contrast.\\
\hline
$\epsilon$ & $\epsilon=\Phi_p/\Phi_s$. Planet-to-star flux ratio.\\
\hline
SNR & Signal-to-noise ratio.\\
\hline
$N_\star$ & Total signal from star (photo-electrons).\\
\hline
$I$ & Spectral irradiance in image plane. (W/m$^2$/$\mu$m)\\
\hline
$E_0$ & Field due to on-axis star in image without aberrations.\\
\hline
$E_\mathrm{speck}$ & Stellar field due to mid-spatial frequency pupil aberrations (i.e. speckles).\\
\hline
$\sigma_\mathrm{speck}$ & Speckle noise.\\
\hline
$S_\mathrm{speck}$ & Effective speckle signal (photo-electrons).\\
\hline
$S_0$ & Effective signal from star without aberrations (photo-electrons).\\
\hline
$C_\mathrm{speck}$ & Raw contrast due to speckles alone.\\
\hline
$C_0$ & Raw contrast without aberrations.\\
\hline
$\boldsymbol{\xi}$ & Spatial frequency vector.\\
\hline
$F\#$ & Focal ratio.\\
\hline
$\omega$ & Wavefront error at critical spatial frequency (waves rms).\\
\hline
$\epsilon_\mathrm{speck,1}$ & Flux ratio limit set by speckles.\\
\hline
$\epsilon_\mathrm{speck,2}$ & Flux ratio limit set by the speckle interference cross term.\\
\hline
$\epsilon_\mathrm{phot}$ & Flux ratio limit set by photon noise.\\
\hline
\end{tabular}
\caption{List of variables used to derived coronagraph metrics.}
\label{table:vars}
\end{table}

The inner working angle and outer working angle of a coronagraph give the range of angular separations over which the system is designed to provide high sensitivity to planets. The inner working angle is either defined as the minimum angular separation at which the total energy throughput reaches 50\% of its maximum value or the raw contrast reaches a threshold value. Similarly, the outer working angle is either defined as the maximum angular separation where throughput is greater than 50\% of its maximum or the raw contrast increases above a threshold. 

It is important to include expected imperfections in the coronagraph performance analysis. All ground- and space-based telescopes have significant, dynamic wavefront errors. Those that change on time scales much faster than the single exposures are known as jitter. Tip-tilt jitter is typically simulated by propagating an ensemble of incoherent point sources in a small region about the optical axis through the instrument, summing the point spread functions, and determining the effective $\eta_p$ and $\eta_s$. The ensemble of incoherent point sources should represent the convolution of the tip-tilt jitter and stellar brightness distributions to also account for the finite size of the star. 

Another key coronagraph design consideration is the sensitivity to low order aberrations. A coronagraph designed to only suppress a plane wave may set unmanageable wavefront error requirements on the telescope and/or AO system to achieve the goal raw contrast. On the other hand, a coronagraph that is more robust to low order aberrations generally has a larger inner working angle, but passively rejects starlight in the presence of certain wavefront errors (depending on the design). A more robust coronagraph relaxes stability requirements on the telescope and AO instrument needed to maintain the desired raw contrast in the presence of AO residuals as well as mechanical and thermal motions in the telescope and instrument. What is more, such coronagraph designs are also less sensitive to differential polarization aberrations, which are generally well described by a linear combination of the few lowest order Zernike polynomials.


The coronagraph masks must be designed to be robust to optical alignment errors and beam magnification. For instance, the pupil magnification uncertainty at the exit pupil of the James Webb Space Telescope (JWST) is $\pm$1-2\%. For the WFIRST CGI, the magnification and translation uncertainties are 0.5\% and 0.2-0.5\% of pupil diameter, respectively. Pupil masks in apodized coronagraphs must therefore oversize input pupil obstructions such as the secondary mirror, its support struts, and/or gaps between mirror segments at the cost of throughput performance. 


\subsection{Planet-to-star flux ratio limits}

In the presence of stellar photon noise, the signal-to-noise ratio for detection is $\mathrm{SNR}=S_p/\sqrt{S_s}.$ For the sake of simplicity in this discussion, we assume $\eta_s$, $\eta_p$, and the planet-to-star flux ratio, $\epsilon=\Phi_p/\Phi_s$, are approximately constant as a function of wavelength. The expression for the planet and star signals simplify to 
\begin{align}
    S_p &= \eta_p \epsilon  N_\star,\\
    S_s &= \eta_s N_\star,
\end{align}
where
\begin{equation}
    N_\star = \int_{\Delta\lambda}  \Phi_s(\lambda) A \Delta t \, q(\lambda) T(\lambda)d\lambda
\end{equation}
is the maximum possible stellar signal in units of photo-electrons. From this point, we assume the $\mathbf{x_0}$ argument is implicit. The stellar photon noise limited SNR becomes
\begin{equation}
    \mathrm{SNR}=\frac{\eta_p}{\sqrt{\eta_s}} \epsilon \sqrt{N_\star}=\epsilon\sqrt{\frac{\eta_p N_\star}{C}}.
\end{equation}
The detection limits of a observation are most often communicated in terms of the minimum detectable planet-to-star flux ratio $\epsilon_\mathrm{lim}$. To achieve an SNR of unity, 
\begin{equation}
    \epsilon_\mathrm{lim}=\sqrt{\frac{C}{\eta_p N_\star}}.
\end{equation}
An exoplanet with $\epsilon=10^{-10}$ may be detected if the $C$ is small enough to ensure that the flux ratio limit set by photon noise $\epsilon_\mathrm{lim}$ is less than $10^{-10}$. A system with $\eta_p=10\%$ and $N_\star\approx10^{12}$, which is typical for a one hour integration on a solar type star at 10 pc, requires $C\approx10^{-10}$.

The dominant noise source in high-contrast images is often spatial speckle noise. Speckles appear as blobs in the image whose spectral irradiance at a single wavelength is described by
\begin{equation}
    I(\mathbf{x},\lambda) = |E_0(\mathbf{x},\lambda) |^2 + |E_\mathrm{speck}(\mathbf{x},\lambda)|^2 + 2\mathrm{Re}\{E_0(\mathbf{x},\lambda)^*E_\mathrm{speck}(\mathbf{x},\lambda) \},
\end{equation}
where $|E_0(\mathbf{x},\lambda) |^2 \approx\text{PSF}_\text{coro}(\mathbf{x},0,\lambda)$ with unaberrated wavefront and $E_\mathrm{speck}(\mathbf{x},\lambda)$ represents the field due to a small additive wavefront error normalized by the coronagraph throughput\cite{Racine1999,Bloemhof2001,Sivaramakrishnan2002,Perrin2003,Aime2004,Soummer2007}. The additive wavefront errors that cause speckles interfere with the underlying residual starlight. 

Aberrations that change slowly may be calibrated using differential imaging techniques, such as Angular Differential Imaging (ADI) \cite{Davies1980,Marois2006}. Those that change on small timescales (e.g. atmospheric residuals) average out into a speckle halo, which may also be subtracted from the image. In these cases, we are only left with the photon noise from the subtracted speckles. However, aberrations that vary on intermediate timescales (i.e. quasi-static) are the most problematic since they are responsible for residual spatial speckle noise after differential imaging. The remaining spatial speckle noise introduces a systematic noise floor for detecting companions through direct imaging.

The SNR when limited by spatial speckle noise is $\mathrm{SNR}=S_p/\sigma_\mathrm{speck}$, where\cite{Aime2004,Soummer2007}
\begin{equation}
    \sigma_\mathrm{speck}^2 \approx  S_\mathrm{speck}^2 + 2S_\mathrm{speck}S_0,
\end{equation}
where $S_\mathrm{speck}$ and $S_0$ are the signals that would be detected from the aberrated and unaberrated wavefronts alone. Expanding the signal terms:
\begin{equation}
    S_\mathrm{speck} \approx C_\mathrm{speck} \eta_p N_\star,
\end{equation}
\begin{equation}
    S_0 \approx C_0 \eta_p N_\star,
\end{equation}
where we have made the approximation that $\eta_p$ is constant over the spectral bandwidth. The raw contrast of a speckle at a position $\mathbf{x}^\prime$ is directly tied to the amplitude of an aberration at spatial frequency $\boldsymbol{\xi}=\mathbf{x}^\prime/\lambda F\#$ cycles per pupil diameter, where $F\#$ is the focal ratio. For a wavefront error $\omega$ at the spatial frequency $\boldsymbol{\xi}$, the raw contrast at $\mathbf{x}$ is $C_\mathrm{speck} = 2(\pi\omega)^2$, where $\omega$ is in units of waves rms\cite{Malbet1995,Ruane2018_JATIS}. $C_0$ is the minimum possible raw contrast (unaberrated case). The SNR may be written 
\begin{equation}
    \mathrm{SNR}=\frac{\epsilon \eta_p  N_\star}{\sqrt{C_\mathrm{speck}^2 \eta_p^2 N_\star^2 + 2C_\mathrm{speck}C_0 \eta_p^2 N_\star^2}}=\frac{ \epsilon }{\sqrt{C_\mathrm{speck}^2  + 2C_\mathrm{speck}C_0}}.
\end{equation}
The planet-to-star flux ratio at which SNR is unity is given by
\begin{equation}
    \epsilon_\mathrm{lim}=\sqrt{C_\mathrm{speck}^2  + 2C_\mathrm{speck}C_0}.
\end{equation}
More generally, we write the flux ratio limits as three terms added in quadrature:
\begin{equation}
    \epsilon_\mathrm{lim}=\sqrt{\epsilon_\mathrm{speck,1}^2  + \epsilon_\mathrm{speck,2}^2+\epsilon_\mathrm{phot}^2}, 
\end{equation}
where $\epsilon_\mathrm{speck,1}=C_\mathrm{speck}$, $\epsilon_\mathrm{speck,2}=\sqrt{2C_\mathrm{speck}C_0}$, and $\epsilon_\mathrm{phot}=\sqrt{C/\eta_p N_\star}$ represent the individual flux ratio limit terms. 

Coronagraph instruments are often designed to minimize the limiting flux ratio. The first term, $\epsilon_\mathrm{speck,1}$, is set by wavefront stability and the effectiveness of the differential imaging strategy. It is minimized by creating a dark hole in the stellar PSF with the adaptive optics system and maintaining it throughout the observation. The second term, $\epsilon_\mathrm{speck,2}$, describes the interaction between wavefront errors and starlight diffracted through the coronagraph in the unaberrated case. Lastly, $\epsilon_\mathrm{phot}$ is the stellar photon noise term. 

Another way to define coronagraph design metrics is to predict and minimize the expected integration time for detection. For example, in the stellar photon noise limited regime, the integration time for SNR=1 is 
\begin{equation}
    \Delta t = \frac{C}{\eta_p}\frac{1}{\epsilon^2\dot{N_\star}}=\frac{\eta_s}{\eta_p^2}\frac{1}{\epsilon^2\dot{N_\star}}, 
\end{equation}
where $\dot{N_\star}=N_\star/\Delta t$ is the count rate. Thus, an optimal coronagraph minimizes $\eta_s/\eta_p^2$. Generalizing the SNR expressions above to include other noise sources, 
\begin{equation}
    \mathrm{SNR} = \frac{S_p}{\sqrt{S_s + \sigma_\mathrm{det}^2+ \sigma_\mathrm{bkgd}^2}},
\end{equation}
where $\sigma_\mathrm{det}$ is the detector noise and $\sigma_\mathrm{bkgd}$ is the background noise (e.g. thermal background). Since the additional noise terms do not depend on the coronagraph throughput, we can characterize the relative importance of each additional noise term as constants $a_n$, where $\sigma_n^2 = a_n N_\star$, and write the integration time as follows:
\begin{equation}
    \Delta t = \frac{1}{\eta_p^2\epsilon^2\dot{N_\star}}\left[ \eta_s + \sum_n a_n\right]. 
\end{equation}
Thus, the coronagraph is only designed to suppress starlight enough such that the stellar photon noise no longer dominates the error budget. 

Finally, when limited by mid-spatial frequency aberrations that generate speckles, an optimal coronagraph minimizes their photon noise contribution by maximizing the coronagraph throughput. We write $\sigma_\mathrm{speck} = b N_\star$, where $b$ is a constant representing the strength of the spatial speckle noise. The expression for the integration time becomes 
\begin{equation}
    \Delta t = \frac{1}{\epsilon^2\dot{N_\star}}\left[\frac{ \eta_s + \sum_n a_n}{\eta_p^2 - b^2/\epsilon^2}\right]. 
\end{equation}
If $b>\eta_p\epsilon$, spatial speckle noise prevents detection (i.e. integration time is negative). Therefore, an optimal coronagraph minimizes $|\Delta t|$, but maintains $\eta_p>b/\epsilon$ to ensure planets may be detected. What is more, the removal of speckles through differential imaging is most efficient when the coronagraph throughput any maximized because the SNR of the speckle measurement is also maximized in each frame.

\subsection{Planet characterization metrics}

Requirements for planet characterization differ from that of planet detection in a number of ways. First, the planet photon noise must be included in the SNR calculation when measuring planet photometry. More importantly, residual spatial speckle noise outside of AP can be ignored. A coronagraph optimized to minimize the photon noise terms is optimal for planet characterization. 

In addition to modifying the noise terms included in the SNR expression above, the optical performance is also improved when characterizing a known directly imaged planet. By focusing the wavefront control degrees of freedom on creating a localized dark hole at the planet position, starlight can be further suppressed. Futhermore, template matching methods can provide gains by defining observables that use high spectral resolution information to better differentiate between planet and starlight and constrain the abundance of molecules in the planet's atmosphere \cite{SparksFord2002,Riaud2007,Snellen2015,Wang2017_HDCI,Mawet2017_HDCII,Lovis2017}.

\subsection{Predicting scientific yield}

Often improving one coronagraph performance metric implies a trade-off with another. To find the optimal balance between the performance metrics listed above, scientific yield calculations can serve as a comprehensive coronagraph performance metric as opposed to optimizing the instrument for a given star or planet type. For example, the requirements of HabEx and LUVOIR are articulated in terms of the number of HZs searched and the number of planets with well determined spectra and/or orbits. Other than the spectral range and resolution, the coronagraph requirements are set by optical designers. The trade studies in that case are carried out using the Altruistic Yield Optimization (AYO) method developed by Stark et al. \cite{Stark2015,Stark2016}.

\section{Coronagraph design elements}

Many coronagraph design studies have been carried out in the context of ground-based imagers (e.g. Palomar/P1640\cite{Hinkley2011}, Gemini/GPI\cite{Macintosh2006}, and VLT/SPHERE\cite{SPHERE2008}) and coronagraph instruments for space telescope concepts (TPF-C\cite{Mouroulis2005}, PICTURE\cite{Mendillo2012,Douglas2015}, Exo-C\cite{exoc}, WFIRST/CGI\cite{Noecker2016}, HabEx\cite{Mennesson2016}, and LUVOIR\cite{Pueyo2017}). Most can be thought of as a subset of the optical layout shown in Fig.~\ref{fig:layout}. Typical coronagraph layouts are listed in Table~\ref{table:existingCoroTypes} with their components and the instruments in which they have been implemented or will be in the future. In this section, we describe the most common coronagraph components and their behaviors. The ultimate goal in coronagraph design is to jointly optimize the degrees of freedom provided by each component to maximize the scientific yield of the instrument. 

\begin{figure}[t]
    \centering
    \includegraphics[width=\linewidth]{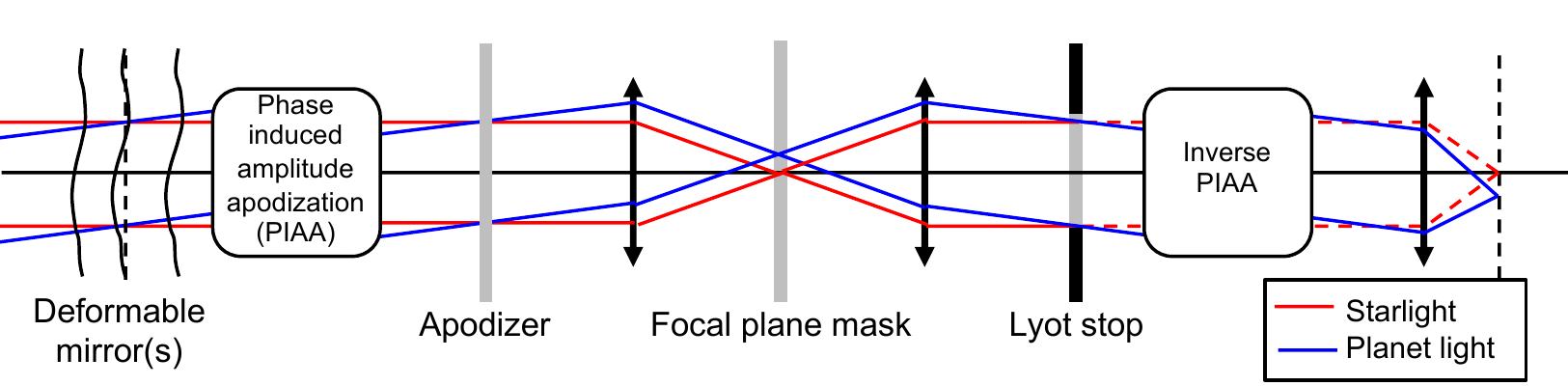}
    \caption{Schematic of a coronagraph with one or more deformable mirrors (DMs), optics for phase induced amplitude apodization (PIAA), an apodizer mask, focal plane mask, Lyot stop, and inverse PIAA optics. The system is optimized to suppress the light from the star (red rays) while maintaining the planet light (blue rays). }
    \label{fig:layout}
\end{figure}



\begin{table}[t!]
\small
\centering
\begin{tabular}{|c|c|c|c|c|c|c|l|} 
\hline
Coronagraph type &\multicolumn{3}{c|}{Pupil apodization} & \multicolumn{2}{c|}{Focal plane mask} & Lyot Stop & Instrument or R\&D bench\\ 
 \cline{2-6}
& Ampl. & Phase & PIAA$^\dagger$ & Ampl. & Phase  & & \\ [0.5ex] 
  \hline
Lyot    &   &  &   &$\times$&        &$\times$ & Ubiquitous \\
\hline
Band-limited Lyot   &   &  &   &$\times$&        &$\times$ & JPL/HCIT \\
\hline
Hybrid Lyot     & $\times$ & $\times$  &$\times$    &$\times$&    $\times$    &$\times$ & JPL/HCIT, WFIRST, Exo-C \\
\hline
Roddier, Dual zone, &   &  & & &$\times$&$\times$&Paris/THD, VLT/SPHERE\\
Four quadrant &   &  & & & &&JWST\\
phase masks&&&&&&&\\
\hline
Vortex &   &  & & &$\times$&$\times$&JPL/HCIT, Caltech/HCST,\\
&&&&&&&Keck/NIRC2,Subaru/SCExAO \\
&&&&&&&LBT/LMIRCam, VLT/NACO,\\
&&&&&&&HabEx \\
\hline
Apodized Vortex & $\times$ & $\times$  & $\times$  &        &$\times$& $\times$ & Caltech/HCST \\
\hline
Shaped pupil & $\times$  & & & & & & Subaru/SCExAO, Princeton/HCIL\\
\hline
Shaped pupil Lyot & $\times$ & & & $\times$ & & $\times$ & WFIRST, LUVOIR\\
\hline
Apodized pupil Lyot & $\times$ & & & $\times$ & & $\times$ & Gemini/GPI, VLT/SPHERE,\\
&&&&&&&STScI/HiCAT, LUVOIR\\
\hline
Apodizing Phase Plate & & $\times$ & & & & & MagAO, Subaru/SCExAO\\
\hline
PIAA$^\dagger$ coronagraph & & & $\times$ & $\times$ & & $\times$ & Subaru/SCExAO, Ames/ACE\\
\hline
PIAACMC$^{\dagger\dagger}$          & & & $\times$ & $\times$ & $\times$ & $\times$ & Subaru/SCExAO, Ames/ACE\\
\hline
VNC$^{\dagger\dagger\dagger}$ & & $\times$& &  &  & $\times$ & PICTURE(-B), Goddard VNT\\
\hline
\end{tabular}
\caption{Common coronagraph architectures, their components, and example instruments in which they have been implemented or are part of a planned design. Pupil apodization may be in the form of amplitude, phase, or phase induced amplitude. Focal plane masks apply amplitude, phase, or both. $^\dagger$Phase induced amplitude apodization. 
$^{\dagger\dagger}$PIAA complex mask coronagraph. $^{\dagger\dagger\dagger}$Visible Nulling Coronagraph. 
}
\label{table:existingCoroTypes}
\end{table}

\subsection{Apodizers}

Apodizers are masks placed in the pupil of an optical system to reduce the diffraction from a point source away from the core of the PSF\cite{Jacquinot1964,Breckinridge1982,Watson1991,Nisenson2001}. The simplest example of an apodization function is a Gaussian function whose Fourier transform in self-similar and has no diffraction rings. However, in real imaging systems, only apodization functions with compact support are possible to implement. 

Early work on apodized pupil Lyot coronagraphs (APLCs) solved for the apodization function using an integral eigenvalue problem, which produced prolate spheroidal functions optimized for circular or annular pupil shapes\cite{Slepian1965,Soummer2003_APLC,Soummer2005_APLC,Soummer2009_APLCII,Soummer2011_APLCIII}. In the APLC concept, the starlight is concentrated behind a circular focal plane mask to prevent it from reaching the final image plane. Later work used more generalized numerical optimization to produce smooth and continuous apodization functions with improved spectral bandwidth, inner working angle, and robustness to low order aberrations\cite{NDiaye2015_APLCIV}. 

Smooth, grayscale apodization functions may be converted into binary amplitude masks by using patterned microdots\cite{Dorrer2007,Guerri2008,Martinez2009a,Martinez2009b,Sivaramakrishnan2009,Sivaramakrishnan2010,Thomas2011,Mawet2014,Zhang2018}. Half-toning, or error diffusion techniques, are effective for reproducing the desired amplitude functions assuming the dots are sufficiently small with respect to the pupil area\cite{FloydSteinberg1976,Ulichney1988}. This approach was used to manufacture the Palomar/P1640\cite{Hinkley2011}, Gemini/GPI\cite{Sivaramakrishnan2010} and VLT/SPHERE\cite{Guerri2008} APLC coronagraphs. 

Simultaneous developments on binary amplitude apodizers used linear programming methods to design shaped pupil apodizers\cite{Kasdin2003}. Shaped pupils are apodizers with a limited outer working angle; that is, they produce a dark hole about the point source with a set outer radius. Making use of powerful optimization tools such as Gurobi\footnote{http://www.gurobi.com/} has extended the use of binary amplitude apodizers to arbitrary telescope pupils\cite{Carlotti2011}. Combining the design methods used for shaped pupils into Lyot coronagraphs have led to promising solutions for centrally obscured apertures on future space telescopes\cite{Zimmerman2016,NDiaye2016_APLCV}. 

Apodized phase plate (APP) coronagraphs\cite{Codona2004,Kenworthy2007,por2017optimal} are similar in principle to shaped pupils, but produce a dark hole in the PSF by only modifying the phase in the pupil. APPs are enabled by the direct writing of vector phase masks in liquid crystal\cite{Miskiewicz2014} and the use of multiple layers to create broadband achromatic phase shifts\cite{Komanduri2013}. A key advantage of the APP approach over shaped pupils is that it is possible to produce a dark hole on one side of the PSF, which results in a more favorable trade-off between inner working angle, throughput, and spectral bandwidth. Splitting the PSF by circular polarization gives two PSFs with the dark hole on either side, allowing for full 360$^\circ$ coverage\cite{Snik2012,Otten2017}. 

In addition to Lyot coronagraphs, apodizers have also been designed for phase mask coronagraphs for obstructed and/or segmented pupils. The shaped pupil approach may be used with finite-radius focal plane masks\cite{Carlotti2013,Carlotti2014}. For focal plane phase masks with infinite radius, the grayscale apodizers needed to account for the central obscuration may be derived analytically\cite{Mawet2013_ringapod,Fogarty2017}. Another approach known as Auxiliary Field Optimization (AFO) can be used to design grayscale apodizers for phase masks coronagraphs on arbitrary pupils\cite{Jewell2017,Ruane2016_SPIE}. 



\subsection{Focal plane masks}


Long after the initial invention of the Lyot coronagraph\cite{Lyot1939}, adaptive optics provided a means to correct and stabilize the wavefront allowing for the introduction of smaller inner working angle coronagraphs. The first evolution of the Lyot coronagraph was the introduction of band-limited\cite{Kuchner2002,Crepp2007} and phase-only focal plane masks that diffract the starlight outside of the Lyot stop. Early phase mask designs were the Roddier-Roddier coronagraph \cite{Roddier1997} and the dual phase mask coronagraph (DZPM)\cite{Soummer2003_DZPM}, which apply phase shifts in circular and annuluar regions near the center of the stellar PSF. 

Another family of masks was introduced with the advantage of having no radial features to improve broadband performance and robustness to tip-tilt errors associated with pointing the star beam at the center of the focal plane. The first effective version was the four quadrant phase mask (FQPM)\cite{Rouan2000,Riaud2001,Lloyd2003} followed by designs with increased the number of sectors \cite{Murakami2008,Murakami2012}. Masks that use continuouse azimuthal phase ramps are the known as vortex coronagraphs, which avoid dead zones between the mask segments\cite{Mawet2005,Foo2005}. It was noted early on that sector-based and vortex coronagraph masks can all be expressed as linear combinations of azimuthal Fourier modes with even topological charge\cite{Jenkins2008}. Achromatic phase masks of these varieties have been fabricated using both scalar-\cite{Swartzlander2008} and vector-based methods (e.g. liquid crystal\cite{Mawet2009} and photonic crystal\cite{Murakami2013} technology). 

More general complex focal plane masks have been developed for coronagraphs on arbitrary apertures using numerical optimization in the context of PIAACMC designs\cite{Guyon2010,Guyon2014,Newman2015,Newman2016,Knight2017} and phase-only focal plane masks\cite{Ruane2015_SPIE}. Current design efforts for WFIRST and future space telescopes combine phase and amplitude in the focal plane using metal and dielectric layers. Among those are the Hybrid Lyot Coronagraph (HLC)\cite{Trauger2016} and recently developed hybrid APLC \cite{Ndiaye2018} designs.


\subsection{Phase induced amplitude apodization (PIAA)}
Phase Induced Amplitude Apodization (PIAA)\cite{Guyon2005} relies on geometrically remapping of the rays in order to redistribute the light and generate an apodization function in the pupil. For the case of Gaussian apodization, this is typically done with two optics: one to refract rays at the edge of the pupil inwards and one to recollimate the beam. PIAA was first practically demonstrated in the laboratory\cite{lozi2009} within an early version of the SCExAO testbed~\cite{Jovanovic2015}. The original tests used two aspheric CaF$_{2}$ lenses which acted to fill in the secondary obstruction. Due to limitations in manufacturing, the field did not reach zero amplitude at the edges of the pupil as per the design. Instead a binary apodizer was used in conjunction with the lenses to eliminate the ringing in the focal plane. 

Given the manufacturing restrictions on creating aspheric surfaces with rapid changes in sag profile, Guyon et al. reworked the original PIAA design to soften the apodization and utilize a complex mask in the focal plane in conjunction with the lenses. This new coronagraph is known as a PIAACMC\cite{Guyon2010} and offers an inner working angle as low as $1~\lambda/D$. It has also been demonstrated through simulation that the PIAACMC can reach similar performance levels on the segmented and obscured apertures and was therefore an early candidate design for the WFIRST/CGI\cite{kern2016piaacmc}. 


\subsection{Active correction with deformable mirrors}

Deformable mirrors (DMs) are used to correct for small wavefront errors that otherwise limit the performance of coronagraphs. The DM is usually controlled using a interaction matrix built to link the movements on the DM actuators to their effect in the image plane using a focal plane wavefront sensing technique. Example wavefront sensing techniques are pair-wise probing \cite{Borde2006, Giveon2011}, the self-coherent camera (SCC)\cite{Mazoyer2014}, and coronagraphic focal-plane wave-front estimation for exoplanet detection (COFFEE)\cite{Paul2013}. Control techniques that aim to minimize the estimated electrical field in the focal plane use matrix inversion with regularization, also known as electric field conjugation (EFC)\cite{Giveon2007}. EFC algorithms have been refined to minimize the actuator strokes\cite{Pueyo2009} and finally to include the use of Kalman filters to estimate the stellar electric field recursively during wavefront correction\cite{Riggs2016}.

Similar techniques may be used to correct for amplitude aberrations with a single DM on a half dark hole \cite{Borde2006} or on a symmetrical dark hole with a symmetrical dark hole with two DMs\cite{Pueyo2009}. These matrix-based techniques typically rely on a linear relationship between the relatively small phase introduced by the DM(s) and its effect in the focal plane. 

A technique called Active Correction for Aperture Discontinuities (ACAD) \cite{Pueyo2013} expanded the formalism of the PIAA coronagraph to correct for these non-axisymmetric discontinuities in the aperture. However, DM correction is limited in both achievable spatial frequencies and in actuator stroke. For this reason, the original ACAD algorithms produced limited results with realistic DMs\cite{Mazoyer2016}. Therefore, the Active Correction for Aperture Discontinuities-Optimized Stroke Minimization (ACAD-OSM) method \cite{Mazoyer2018a} adopted a new approach, also based on an interaction matrix and a focal plane wavefront sensor technique to obtain higher contrast levels. To overcome the limitation of the small-stroke assumption, the interaction matrix is frequently re-computed using a new initial state. The strokes introduced with each matrix remain small but the final DM shape corrects for discontinuities in the aperture. A similar technique (using re-computation of the matrix) is also used by the EFC-based Fast Linearized Coronagraph Optimizer (FALCO)\cite{Riggs2018} as well as the hybrid Lyot coronagraph mode of WFIRST\cite{Krist2015,Trauger2016}. The Auxiliary field optimization (AFO) \cite{Jewell2017} technique uses an expectation-maximization algorithm to compute the DM shapes. 

The main advantage of active techniques using DMs is that they may be used to correct simultaneously for known static aberrations, diffraction from struts supporting the secondary mirror and gaps between segments in the primary mirror, as well as evolving and unknown aberrations, such as phase errors due to the atmosphere, segment phasing \cite{Mazoyer2018a}, or amplitude aberrations due to misalignment of the optics \cite{Mazoyer2017}. Performance in terms of raw contrast and throughput depends on the DM arrangement, particularly the size of the DMs and distance between them\cite{Beaulieu2017,Mazoyer2018b}.

All of the active techniques described here are compatible with most coronagraphs layouts described above. Since DMs are limited in number of actuators and strokes, some diffraction will preferentially be suppressed with fixed optics (e.g. apodizers), while others, especially those expected to vary over time, will need to be corrected with DMs. 

\subsection{Lyot stops}

The invention of the coronagraph as we use it today was marked by Lyot's introduction of an iris downstream of the opaque focal plane mask to suppress diffraction rings in the image plane\cite{Lyot1939}. Now, the Lyot stop is a critical component of all coronagraphs that make use of a focal plane mask. In many coronagraph designs, the main suppression mechanism is to relocated starlight outside of the geometric pupil after the focal plane mask where it is blocked by a specially designed Lyot stop. A typical Lyot stop resembles the re-imaged pupil, with oversized boundaries to account for misalignment and optical distortion in the instrument. Only a few studies have explored optimizing the Lyot stop numerically\cite{Zimmerman2016,Ruane2015_LPM}. Apodizers in the plane of the Lyot stop may also be designed to help suppress on-axis starlight\cite{Ruane2015_LPM} as well as diffraction from an off-axis stellar companion\cite{Kaufl2018}.  

\section{Publicly available coronagraph design tools}

\subsection{PROPER: Optical propagation library}
PROPER\footnote{\href{http://proper-library.sourceforge.net/}{http://proper-library.sourceforge.net/}} is an optical wave propagation library for simulating an optical system using Fourier transform algorithms, namely Fresnel propagation and angular spectrum methods. It is currently available for IDL (Interactive Data Language), Matlab, and Python (2.7 \& 3.x). It includes routines to create complex apertures, aberrated wavefronts, and deformable mirrors\cite{Krist2007,KristTDEM}. PROPER is especially useful for the simulation of high contrast imaging telescopes with coronagraphs. It is distributed as source code, is well documented with a detailed manual, and is relatively easy to use.

\subsection{POPPY: Physical Optics Propagation in Python}
POPPY is an object-oriented Python module for simulating diffraction effects\footnote{\href{https://github.com/mperrin/poppy}{https://github.com/mperrin/poppy}}. 
POPPY was developed as the wavefront propagator for the WebbPSF module\cite{Perrin2012} in an Astropy\cite{the_astropy_collaboration_astropy_2013} compatible framework.
WebbPSF performs system level diffraction simulations of point-spread-functions from the James Webb Space Telescope, including coronagraph masks, in the Fraunhofer domain.
POPPY defines convenient wavefront and optical system classes and supports simultaneous diffraction calculations using multiple processors. 
Additional performance gains over a pure-python implementation are provided by a variety of high-performance numerical  libraries, including graphics processing units\cite{Douglas2018}.
Recent work has extended POPPY to the Fresnel domain, which has enabled plane-to-plane simulation of near-field effects\cite{lawrence_optical_1992} for both space and ground-based coronagraphic observatories, e.g. the Magellan Extreme Adaptive optics system (MagAO-X\cite{Lumbres2018}).


\subsection{HCIPy: High Contrast Imaging for Python}

HCIPy\cite{por2018hcipy} is an object-oriented framework written in Python for performing end-to-end simulations of high-contrast imaging instruments. It relies on the concept of Fields, which unify both the sampling of space and the values at those points. The ubiquitous usage of Fields throughout the whole library makes writing code less error prone. Furthermore, most sampling calculations are handled seamlessly in the background by the library allowing the user to focus on the high-level structure of their own code.

The library defines wavefronts and optical elements for defining an optical system. It provides both Fraunhofer and Fresnel diffraction propagators. Polarization is supported using Jones calculus, with polarizers and waveplates. It implements atmospheric turbulence using thin infinitely-long phase screens and can model scintillation using Fresnel propagation between individual layers. Many wavefront sensors are implemented including a Shack-Hartmann and Pyramid wavefront sensor. Implemented coronagraphs include the vortex, Lyot, and APP coronagraphs. Methods are available for globally optimizing the pupil-plane of any coronagraph, both in phase and amplitude. This code is based on linear optimization\cite{Carlotti2013,por2017optimal}. Coronagraphs that belong to this category include APLCs, APP coronagraphs, shaped pupil coronagraphs, apodized vortex coronagraphs and more.

By including simulation of both adaptive optics and coronagraphy in the same framework, HCIPy allows simulations including feedback from post-coronagraphic focal-plane wavefront sensors to the AO system. HCIPy is available as open-source software\footnote{\href{https://github.com/ehpor/hcipy}{https://github.com/ehpor/hcipy}}. Community input and contributions are welcome.



\subsection{STScI SCDA toolkit}
The SCDA (Segmented Coronagraph Design \& Analysis) research team at the Space Telescope Science Institute (STScI) developed a corongraph design survey toolkit\cite{Zimmerman2016b} with the support of the NASA Exoplanet Exploration Program (ExEP) to explore apodized/shaped pupil Lyot coronagraph solutions for segmented telescope pupils. For each set of design survey parameters, the toolkit generates a pool of AMPL \cite{Fourer1990} linear program scripts to then be executed in queued batches on a computing cluster. These linear programs rely on the Gurobi\cite{gurobi} solver to determine the apodizer mask solution with maximum off-axis transmission for a given set of design constraints (namely the raw contrast goal, dark zone extent and spectral bandwidth, telescope pupil, occulting mask, IWA, OWA, and Lyot stop profile). The object-oriented approach of the SCDA toolkit simplifies the interface for sampling large parameter spaces, and enables flexibility for implementing various mask architectures and symmetry cases. The core module, example notebooks, and documentation are publicly hosted\footnote{\href{https://github.com/spacetelescope/SCDA}{https://github.com/spacetelescope/SCDA}}. 
A detailed algebraic description of the linear Lyot coronagraph propagation models built into the toolkit is given in the appendix of Zimmerman et al. (2016)\cite{Zimmerman2016}.

The original version of the SCDA toolkit used the AMPL language to encode the optimization problem, but recent effort has directly configured the problem in matrix form in Python, thereby speeding up the execution. In this implementation, the Gurobi solver is directly called from Python using the gurobipy package. This code implementation is currently privately hosted at \footnote{\href{https://github.com/astromam/Coronagraphs}{https://github.com/astromam/Coronagraphs}} but it will be made public by the end of 2018. Work is ongoing to integrate this new implementation into the SCDA toolkit and the first applications are presented in these proceedings \cite{fogarty2018}.

\subsection{COFFEE: Coronagraph Optimization For Fast Exoplanets Exploration}

COFFEE is a diffraction-based PIAACMC/APLCMC simulation and optimization software package written in the C programming language. COFFEE source code\footnote{\href{https://github.com/coffee-org/}{https://github.com/coffee-org/}} is organized into modules capable of performing image-based tasks in shared memory; many of these tasks are used during the coronagraph optimization process. COFFEE uses multi-threaded processing and the high-performance parallel computing libraries CUDA and MAGMA for matrix linear algebra computations. To simulate and optimize coronagraphs, the user interfaces with COFFEE through bash scripting to a custom command line interface, calling a sequence of commands to execute each step of the optimization routine. 

The coronagraph design exists as an unfolded configuration in collimated space. The multi-step optimization approach first simulates a monochromatic PIAACMC using an ideal, non-physical focal plane mask while assuming a centrally obscured circular telescope pupil. In subsequent steps the input telescope pupil replaces the centrally obscured telescope pupil so that each coronagraph component can be individually tuned to improve raw contrast and throughput performance. Briefly, the optimization process from an ideal PIAACMC proceeds as follows: the Lyot stop shapes and locations are optimized by Fresnel propagating through a search range along the optical axis from the Lyot plane and applying thresholds at each position to determine where the Lyot stop has the most significant impact. Then the aspheric PIAA shapes are fitted with 2D Fourier and radial cosine basis modes and the ideal focal plane mask transmission is adjusted, both to improve performance given the new Lyot stop location. This process is repeated a number of times to create a high-performance monochromatic PIAACMC for the input telescope pupil. To achieve polychromatic performance with this pupil, the ideal focal plane mask is split into a number of phase-shifting zones of some material, each zone serving as a free parameter in a method of steepest descent algorithm to minimize star light in the coronagraphic point spread function. Once a target contrast value is reported, or the search time expires, the optimization process is complete.


\begin{figure}[t!]
    \centering
    \includegraphics[width=0.8\linewidth,trim={0 0mm 0 0mm},clip]{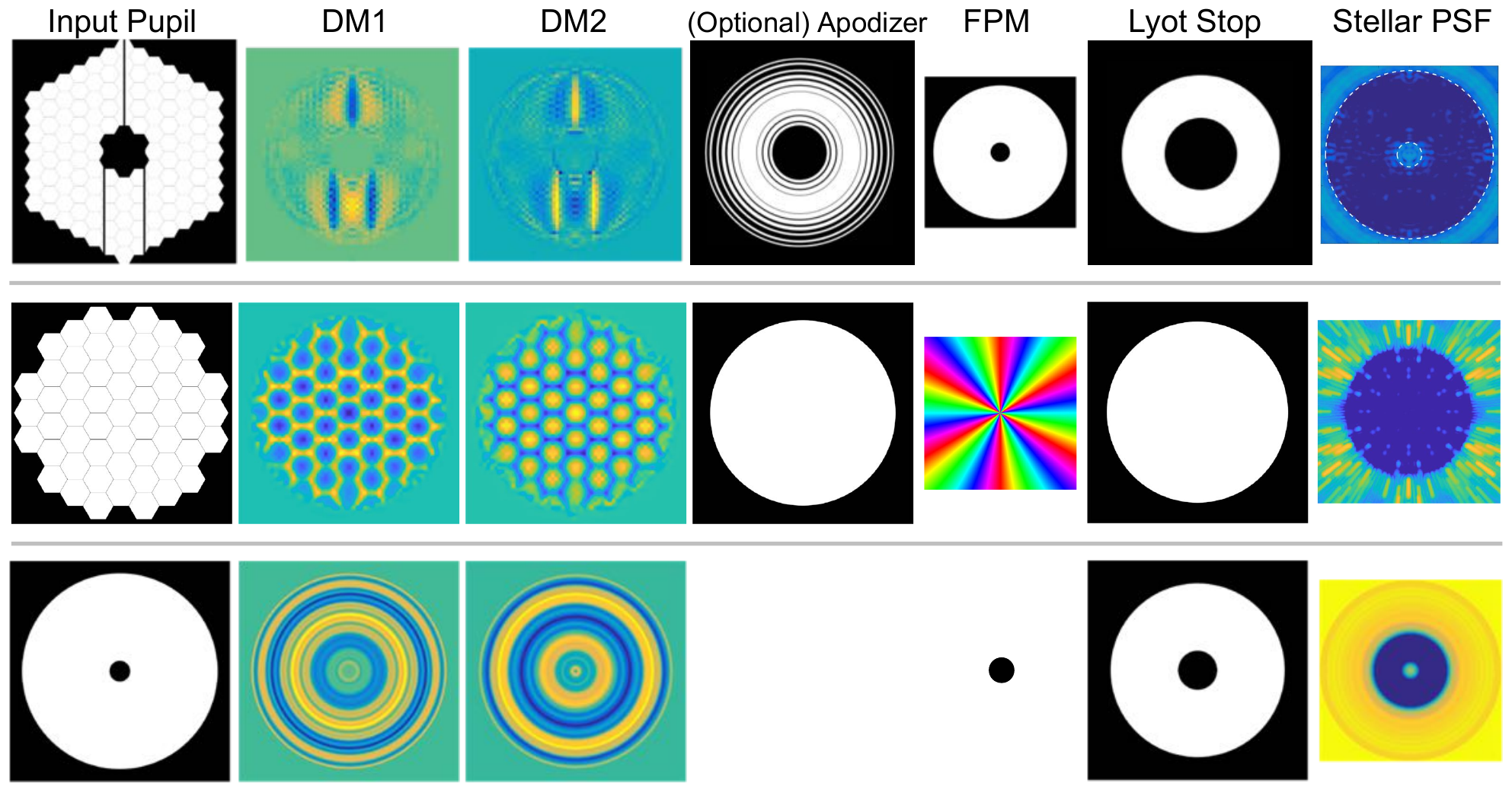}
    \caption{Example coronagraph designs from the FALCO software\cite{Riggs2018}. Each design achieves raw contrast $\leq10^{-10}$. \textbf{First~row}: A 2DM (64$\times$64 actuators each) apodized Lyot coronagraph for a potential LUVOIR aperture\cite{Pueyo2017}. The dark hole is generated over 3.8-27.0 $\lambda/D$ in a 10\% bandwidth.  \textbf{Second~row}: a 20\% BW, 2DM (64$\times$64) vortex design from FALCO for the unobscured, segmented aperture. \textbf{Third~row}: Lyot coronagraph with optimized DM shapes that gives a dark hole from 2.4-10$\lambda/D$.}
    \label{fig:falco}
\end{figure}

\subsection{FALCO: Fast Linearized Coronagraph Optimizer}

Fast Linearized Coronagraph Optimizer (FALCO)\cite{Riggs2018} is a freely available\footnote{\href{https://github.com/ajeldorado/falco-matlab}{https://github.com/ajeldorado/falco-matlab}} MATLAB-based code for wavefront sensing and control with the coronagraph types listed in Table \ref{table:existingCoroTypes} and combinations thereof. FALCO makes use of the PROPER routines to simulate the DMs, optical surfaces, and most optical propagation steps. The DM shapes are optimized for any coronagraph using electric field conjugation\cite{Giveon2011}. Figure \ref{fig:falco} shows example solutions for shaped pupil Lyot, vortex, and conventional Lyot coronagraphs. A key feature of FALCO is a 10-100$\times$ reduction in the calculation time of linearized DM response matrices over the typical approach with large, padded FFTs. These speed gains are obtained without loss of accuracy via semi-analytical methods (e.g., Babinet's principle) and sub-windowed propagation in the region of influence for each DM actuator (only a fraction of the full beam is needed). FALCO's efficiency and generality enables comprehensive DM-integrated coronagraph design surveys.



\section{Summary}

We have reviewed optical performance metrics, components, and open-source software available for finding optimal instrument designs for imaging exoplanets with stellar coronagraphs. This document is the first of a series of three papers summarizing the outcomes of the Optimal Optical Coronagraph (OOC) Workshop at the Lorentz Center in September 2017 in Leiden, the Netherlands.

\acknowledgments  

The authors would like to acknowledge the Lorentz Center for hosting and to a large extent funding the Optimal Optical Coronagraph workshop held September 25-29, 2017 at the Lorentz Center in Leiden, the Netherlands. Additional funding for the workshop was provided by the European Research Council under ERC Starting Grant agreement 678194 (FALCONER) granted to Frans Snik. This formed the platform where this work was carried out. G.~Ruane is supported by an NSF Astronomy and Astrophysics Postdoctoral Fellowship under award AST-1602444.

\small
\bibliography{Library}   
\bibliographystyle{spiebib}   

\end{document}